%% file: main.tex
\documentclass[sigconf,balance=false]{acmart}
\usepackage{popets}

\usepackage{enumitem}
\usepackage{fancyhdr}
\usepackage{float}
\usepackage{graphicx}
\usepackage{listings}
\usepackage{subcaption}
\usepackage{xspace}

\setcopyright{popets}
\copyrightyear{YYYY}
\acmYear{YYYY}
\acmVolume{YYYY}
\acmNumber{X}
\acmDOI{XXXXXXX.XXXXXXX}
\acmISBN{}
\acmConference{Proceedings on Privacy Enhancing Technologies}
\newcommand{\water}{\textit{WATER}\xspace}
\newcommand{\waterfullname}{WebAssembly Transport Executables Runtime}

\makeatletter
\newcommand\paragraphwithoutperiod[1]{%
    {\itshape #1}\@ifnextchar\par{\@gobble}{}\ %
}
\makeatother

\begin{document}

\title[Just add \water]{{Just add \water}: WebAssembly-based Circumvention Transports}

\author{Erik Chi}
\affiliation{%
  \institution{University of Michigan}
  \city{}
  \state{}
  \country{}
}
\email{ziyunchi@umich.edu}

\author{Gaukas Wang}
\orcid{0000-0002-9552-0978}
\affiliation{%
  \institution{University of Colorado Boulder}
  \city{}
  \state{}
  \country{}
}
\email{Gaukas.Wang@colorado.edu}

\author{J. Alex Halderman}
\affiliation{%
  \institution{University of Michigan}
  \city{}
  \state{}
  \country{}
}
\email{jhalderm@umich.edu}

\author{Eric Wustrow}
\affiliation{%
  \institution{University of Colorado Boulder}
  \city{}
  \state{}
  \country{}
}
\email{ewust@colorado.edu}

\author{Jack Wampler}
\orcid{0009-0005-8785-0570}
\affiliation{%
  \institution{University of Colorado Boulder}
  \city{}
  \state{}
  \country{}
}
\email{Jack.Wampler@colorado.edu}

\renewcommand{\shortauthors}{Chi et al.}

\input{figure}
\input{table}

\input{0abstract}
\input{1intro}
\input{2relatedworks}
\input{3design}
\input{4impl}
\input{5eval}
\input{6discussion}

\bibliographystyle{ACM-Reference-Format}
\bibliography{ref}

\input{Appendix}
\input{API}

\end{document}

%% file: figure.tex
\newcommand{\FigOverview}{\begin{figure}[ht]
    \centering
    \includegraphics[width=\linewidth]{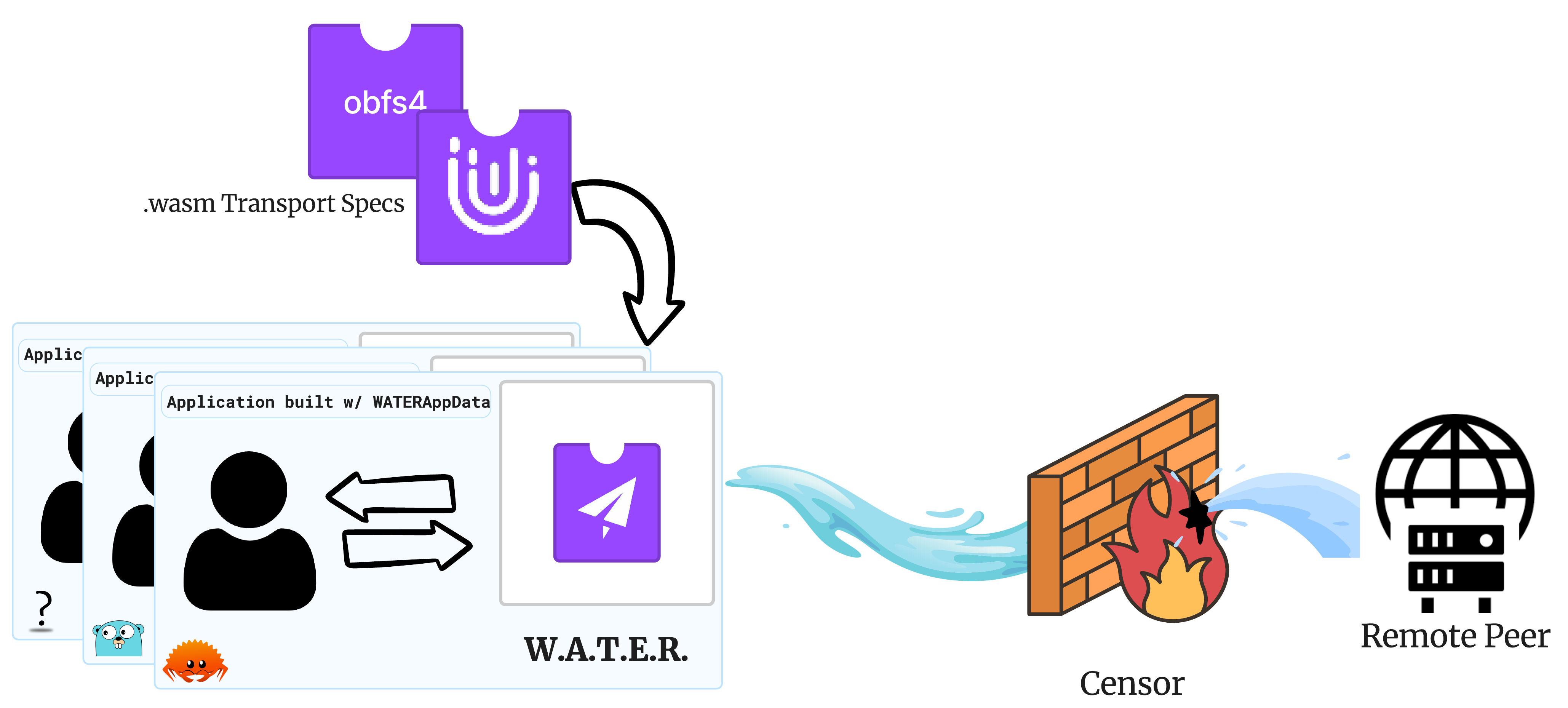}
    \caption{The overview of \water's role in action. With transport specs
    defined by \texttt{.wasm} files distributed out-of-band, \water can
    efficiently switch between transports to use.}
    \label{fig:overview}
    \Description{An overview of how water works}
\end{figure}}

\newcommand{\FigWorkingFlow}{\begin{figure}[t]
    \centering
    \includegraphics[width=\linewidth]{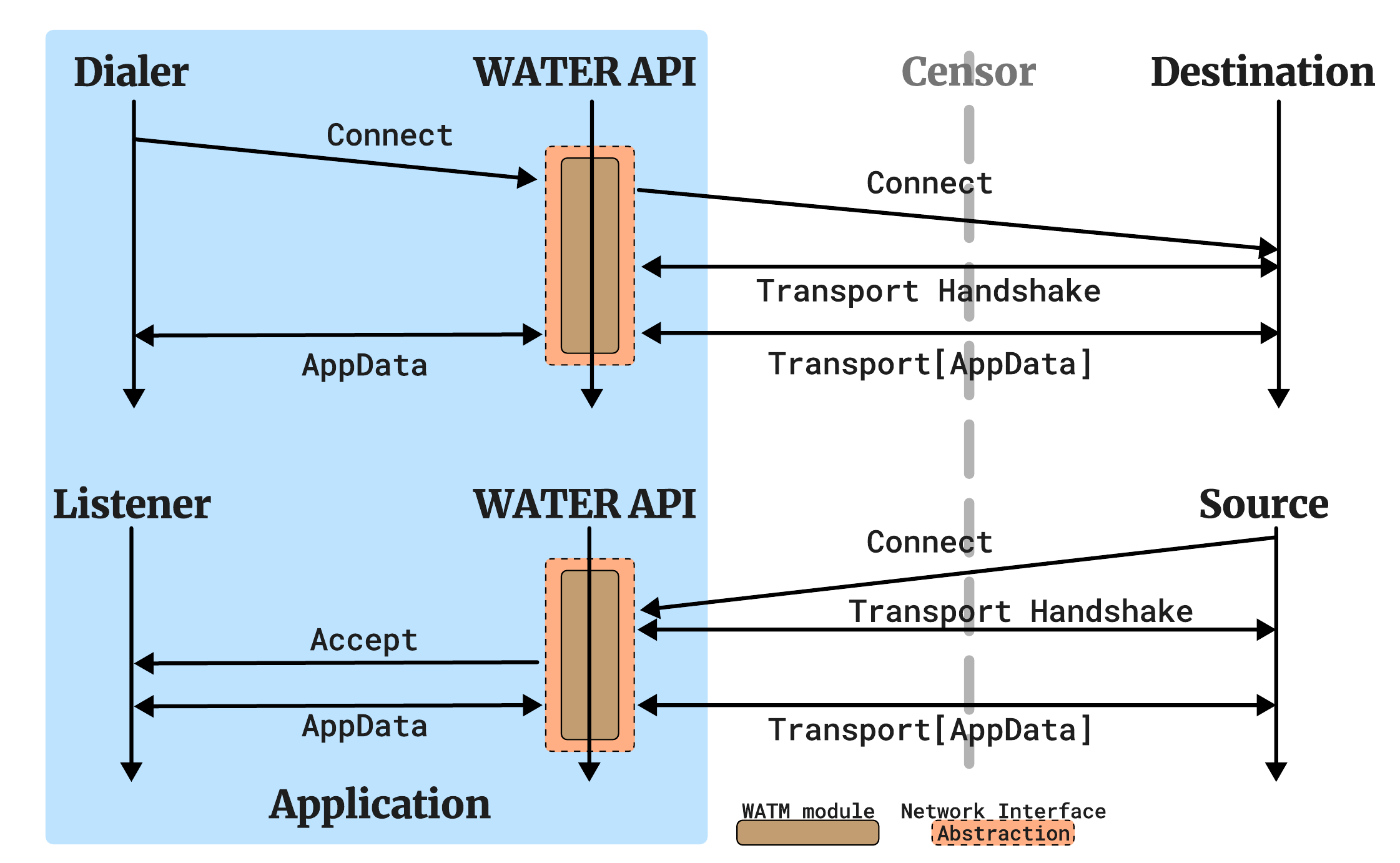}
    \caption{Example connection establishment flows of traditional client
    (\textit{Dialer}) and server(\textit{Listener}) each using a \water
    transport. The dialer actively connects to a remote host upon request by
    caller, with the \water network interface internally managing sockets and IO
    allowing the WATM to transform the byte stream. Similarly, the listener accept the
    incoming connections, allowing the WATM to attempt a handshake with the
    remote host before firing an \textit{accept} hook passing the plaintext
    end to an upstream handler.}
    \label{fig:working-flow}
    \Description{A picture describing workflows of different working models of \water}
\end{figure}}

%% file: table.tex
\newcommand{\TblShadowsocksIperfMac}{\begin{table}[ht]
    \centering
    \begin{tabular}{lcc}
        \toprule
        Travel through & iperf3 - 10s & iperf3 - 600s \\
        \midrule
        shadowsocks-rust & 415 / 411 & 418 / 418 \\
        \textbf{\water-SS} & \textbf{56.5 / 56.5} & \textbf{56 / 56} \\
        Proteus-SS & 96.6 / 83.5 & 68 / 67.8 \\
        \bottomrule
    \end{tabular}
    \caption{Benchmark for General Use Case: Sender / Receiver Throughput (Mb/s) on a MacBook Pro}
    \label{tab:ss_performance_macos} 
\end{table}}

\newcommand{\TblLatencyThroughputCloudlabtopologyFinal}{\begin{table}[ht]
    \centering
    \begin{tabular}{lcc}
        \toprule
        Travel through & Latency & Throughput \\
        \midrule
        shadowsocks-rust (Baseline) & 116us & 2310 Mbps \\
        \textbf{shadowsocks-\water} & \textbf{+493us} & \textbf{2.0\% (46.2 Mbps)} \\ 
        shadowsocks-Proteus & +873us & 4.8\% (110 Mbps) \\
        \midrule
        Raw TCP (Baseline) & 26us & 2210 Mbps \\
        \textbf{plain-\water}  & \textbf{+356us} & \textbf{82.8\% (1830 Mbps)} \\ 
        plain-Proteus & +250us & 105.4\% (2330 Mbps) \\
        \bottomrule
    \end{tabular}
    \caption{Latency \& Throughput benchmark result comparing to the baseline data, 
    with \texttt{msg\_size=512}. Worth noting that an implementation may run 
    slow enough to combine multiple messages into one single \texttt{send()} and 
    achieve better throughput than baseline due to Nagle's algorithm, with the latter 
    has \texttt{TCP\_NODELAY} enabled and cannot combine messages.}
    \label{tab:ss_performance_cloudlab_simulated}. 
\end{table}}

\newcommand{\TblWaterPlainLatencyThroughputCompCloudlabtopology}{\begin{table}[ht]
    \centering
    \resizebox{\columnwidth}{!}{
        \begin{tabular}{lccc}
            \toprule
            P Size(B) & Raw TCP (Baseline) & \water & Proteus \\
            \midrule
            1    & 24us / 6Mbps       & +354us / 166.7\%       & +240us / 183.3\% \\
            64   & 25us / 337Mbps     & +358us / 99.1\%        & +241us / 104.2\% \\
            128  & 25us / 656Mbps     & +341us / 101.4\%       & +241us / 102.3\% \\
            256  & 24us / 1240Mbps    & +358us / 102.4\%       & +242us / 100.0\% \\
            512  & 26us / 2210Mbps    & +356us / 82.8\%        & +250us / 105.4\% \\
            768  & 25us / 3200Mbps    & +358us / 62.5\%        & +250us / 97.8\% \\
            1024 & 26us / 3930Mbps    & +359us / 52.4\%        & +251us / 101.3\% \\
            2048 & 51us / 6390Mbps    & +339us / 31.1\%        & +288us / 88.7\% \\
            4096 & 54us / 9770Mbps    & +334us / 19.2\%        & +292us / 57.8\% \\
            \bottomrule
        \end{tabular}
    }
    \caption{Plain-Relay latency/throughput - CloudLab topology}
    \label{tab:plain_latency_throughput_cloudlab_simulated}
    \end{table}
}

\newcommand{\TblSSPatchCodeDiff}{
    \begin{table}[ht]
        \centering
        \begin{tabular}{|l|l|}
            \hline
            \textit{shadowsocks-rust}\_diff.txt \textbf{(98\%)} & \water-SS\_diff.txt \textbf{(99\%)} \\
            \hline
            1-5     & 1-5     \\
            \hline
            8-196   & 6-194   \\
            \hline
            198-319 & 195-316 \\
            \hline
        \end{tabular}
        \caption{Matched lines from running the \texttt{diff} command on patches between official-SS and \water-SS}
        \label{tab:ss-patch-compare}
    \end{table}
}

\newcommand{\TblShadowsocksCodeCompare}{\begin{table}[ht]
    \centering
    \begin{tabular}{|l|l|}
        \hline
        shadowsocks-rust \textbf{(32\%)} & \water\_SS \textbf{(85\%)} \\
        \hline
        168-242   & 80-154    \\
        \hline
        \dots & \dots \\
        \hline
        2548-2687 & 810-922   \\
        \hline
    \end{tabular}
    \caption{Matched lines in the \water-shadowsocks implementation compares to the official shadowsocks-rust}
    \label{tab:ss-codechangecompare}
\end{table}}

%% file: 0abstract.tex
\begin{abstract}
    
        As Internet censors rapidly evolve new blocking techniques, circumvention tools
        must also adapt and roll out new strategies to remain unblocked.
        But new strategies can be time consuming for circumventors to
        develop and deploy, and usually an update to one tool often requires
        significant additional effort to be ported to others. Moreover, distributing
        the updated application across different platforms poses its own set of
        challenges.
    
        In this paper, we introduce \water (\waterfullname), a novel design
        that enables applications to use a WebAssembly-based application-layer
        (e.g., TLS) to wrap network connections and provide network transports.
        Deploying a new circumvention technique with \water only requires
        distributing the WebAssembly Transport Module(\emph{WATM}) binary and any
        transport-specific configuration, allowing dynamic transport updates without
        any change to the application itself. WATMs are also designed to be generic
        such that different applications using \water can use the same WATM to
        rapidly deploy successful circumvention techniques to their own users,
        facilitating rapid interoperability between independent circumvention tools.
    
    \end{abstract}
    
    \keywords{censorship, circumvention, transport, network, {WebAssembly}}
    
    \maketitle

%% file: 1intro.tex
\section{Introduction}

The arms race between censors and circumventors continues to evolve with new
tools and tactics emerging from both sides: Censors deploy new mechanisms that
block proxies, and in response circumventors develop new techniques that get
around the
blocking~\cite{gfwreport-fully_encrypted,gfwreport-shadowsocks,xue-openvpn,wang-shadowtls}.

Because of its dynamic nature, successful circumvention tools must continually
develop and deploy new strategies and techniques to get around emerging
censorship. For instance, in 2012, Iran blocked several proxies, including Tor,
by using an early form of SSL fingerprinting~\cite{phobos-iran_tor}.
In response, Tor developed obfsproxy~\cite{tor-obfsproxy}, which encrypts all of
its traffic including protocol headers in an attempt to evade protocol
fingerprinting attacks~\cite{frolov-tlsfingerprint,houmansadr-parrot}. While
successful in the short term, this was again insufficient as in censors such as 
China deployed \emph{active probing} attacks to detect early versions of the
protocol~\cite{tor-gfw_obfs2,ensafi-examining}, which prompted circumventors to
develop probe-resistant proxies~\cite{winter-scramblesuit,obfs4}. Censors then
found and exploited other side-channels and vulnerabilities to differentiate
circumvention traffic~\cite{gfwreport-shadowsocks,frolov-detecting}. Once these
problems were addressed, censors began using other features to detect and block
fully-encrypted proxies such as entropy
measurements~\cite{gfwreport-fully_encrypted}, and circumventors responded by
using prefixes that fool these
measurements~\cite{outline-prefix,gfwreport-fully_encrypted}.

Discovering, implementing, and operationalizing circumvention techniques like
these can be burdensome, requiring new code and configurations to be written,
packaged, approved for distribution by app stores, and pushed to users.
Furthermore, each circumvention tool may need to write and maintain their own
version specific to their environment, potentially built using an entirely
different programming language, adding to the cost.

\FigOverview

In this paper, we introduce an approach to ease the burden of developing and
deploying new circumvention techniques in the ongoing censorship arms race. Our
technique uses WebAssembly, a binary instruction format with 
runtime support across various platforms, including web browsers and mobile devices. 
WebAssembly programs can be written in high-level languages such as C or Rust, and 
compiled into a universal binary that runs on any platform with just a WebAssembly runtime. 
We extend their use cases with WebAssembly System Interface~\cite{wasi} (WASI) to allow 
such compiled binaries to make low-level system calls related to network sockets and 
perform I/O operations by defining an experimental interface for WebAssembly Transport Module (WATM), 
which will be described in detail in Section~\ref{sec:watm-design}.
We also create \water, a runtime library allowing circumvention tools to 
use portable circumvention techniques from WATMs.

To use \water, a circumvention tool integrates our \water library into their
client-side program or application. Then, circumventors can build universal 
WATMs that implement new circumvention techniques. These WATMs 
can then be distributed to users over existing data channels, avoiding 
the need to update the whole app, which involves censored mobile app stores or 
dealing with blocked CDNs. One use case can be a technique that implements 
a fully encrypted proxy, written once and distributed to a myriad of circumvention 
tools as a WATM, despite the tools using different software languages and libraries.

\water has a distinct advantage over prior approaches that provide similar
flexibility, such as Proteus~\cite{wails-proteus} or Pluggable
Transports~\cite{pluggable_transports}. Where \water leverages WebAssembly,
new techniques can be written in one of several (and growing~\cite{languages-to-wasm}) 
high-level languages. In contrast, Proteus requires that techniques be written in a 
bespoke domain-specific language (DSL) which is incompatible with the import of other 
code or libraries, and thus must be entirely self-contained. Meanwhile, Pluggable Transports 
must maintain separate, language-specific APIs for each supported programming language 
(currently Go, Java, and Swift), which are incompatible with each other. In contrast, WATMs 
can run in any WATER runtime, currently featuring off-the-shelf implementation in Go and Rust
with the potential of being implemented with any WebAssembly runtime with WASI support. 
Furthermore, these WATMs can be compiled from multiple languages, including popular ones such 
as Rust, Go, Python (with CPython), C (using wasi-libc)~\cite{languages-to-wasm}. \water is also 
well-positioned to benefit from future WASI developments as the standard becomes more widely 
supported and feature-rich.

In the remainder of this paper, we describe our design of \water, implement 
several proof-of-concept WATMs, and compare their flexibility and
performance to existing tools.

%% file: 2relatedworks.tex
\section{Related Work}

Several prior works have focused on providing \emph{protocol agility} to
circumvention tools. We detail these and describe their differences to \water
below.

\paragraphwithoutperiod{(Tor's) Pluggable Transports}~\cite{pluggable_transports} offers a
standardized interface that different tools can integrate into. This allows
circumvention tools (e.g. Tor) that implement the pluggable transport
specification to easily add code for new circumvention transports at compile
time. However, Pluggable Transport interfaces are language-specific, and
currently only Go, Java, and Swift are
supported~\cite{pluggable_transports_github}, making it difficult to create
cross-platform transports that work in multiple projects written in different
languages.

\paragraphwithoutperiod{Proteus} implements an alternate method of dynamically deploying new
transports. It compiles text-based protocol specification files (PSF) and
executes them at low-level with Rust, improving the flexibility in deployment
without losing too much performance~\cite{wails-proteus}. However, Proteus
requires the use of a DSL that forces developers to use a Rust-based syntax and
adopt a restrictive programming style when developing a PSF. This prevents the
direct incorporation of existing tools and increases the difficulty of designing
transports from scratch. Also, as Proteus is implemented in Rust, it is
challenging to integrate Proteus into projects in other programming languages.

\paragraphwithoutperiod{Marionette} is a configurable network traffic obfuscation system used
to counter censorship based on DPI~\cite{dyer2015marionette}. It uses text-based
message templates to apply format-transforming-encryption (FTE) to encode client
traffic into benign looking packets. The templates are constructed using
a DSL along with some customizable encoding and encryption operations. The DSL
is not Turing-complete and is relatively restrictive. To support more complex
protocols, Marionette provides an interface for plugins which requires some
degree of recompilation and therefore still requires redeployment.

%% file: 3design.tex
\section{Design}
\label{sec:design}

There are two key components in \water:
1) a runtime library to be integrated into a circumvention tool to run WATMs and
2) a WebAssembly Transport Module (WATM) as the hot-swappable \texttt{.wasm} binary 
implementing a particular circumvention strategy or transport encoding technique.

\subsection{\water Runtime Library}
\label{sec:core-design}
The runtime library is designed to be easy to integrate into circumvention tools,
allowing them to run hot-swappable WATMs that implement different circumvention
strategies. The runtime library includes a WebAssembly runtime environment to 
execute the WATMs and also presents a standard high-level interface for the 
integrating circumvention client to use any WATM without requiring any knowledge 
about WebAssembly. This allows the client to make calls to the WATM regardless of 
the underlying logic and for the WATM to be able to interact with external resources
such as network sockets, logging, etc.

As depicted in Figure~\ref{fig:working-flow} --- when the client calls
\texttt{\_water\_dial()} in the runtime library, the WATM is launched,
and the WATM-defined \textit{connect} method is invoked. This method may choose to
make one or more TCP connections, which it does by calling \texttt{dial\_host()} \textit{imported} from the
runtime library. Then, the runtime library returns a virtual socket to the
client. When the client writes to the virtual socket, the runtime library passes
the data into a WATM-defined data encoding method, which can transform the data 
based on the protocol and send it out through corresponding network connections. Similarly, 
when the client reads from the virtual socket, it does so through the runtime library and 
WATM-defined data decoding method, allowing the WATM to transform received data from the network. 
In short, the WATM can make arbitrary transformations to the data sent by the client or received 
from the network in order to implement any circumvention technique. A step-by-step workflow of dialing a connection 
is shown in Appendix \ref{sec:workflow-detailed}.

\water also supports server-side connections, by similarly implementing
corresponding \texttt{listen} and \texttt{accept}. 

\FigWorkingFlow

\subsection{WebAssembly Transport Module (WATM)}
\label{sec:watm-design}

A WebAssembly Transport Module (WATM) is a program compiled into a WebAssembly
binary that implements a specific set of expected functions that its Host may invoke, allowing the
\water runtime to interact in a consistent manner while allowing interchangeable
WATMs have the flexibility to apply arbitrary transformations to network traffic.
For example, WATMs could wrap a stream in TLS (implementing TLS within
the WASM binary), could add reliability layers (e.g.
TurboTunnel~\cite{fifield-turbotunnel}, or could arbitrarily shape traffic by changing
the timing or size of packets sent.
The set of exposed functions provided by the \water runtime library allows
the WATM to interact with the network interface abstraction and the runtime
core to manage things like configuration, sockets, cancellation, and logging.

As Rust is one of the most mature languages for writing WATMs,
we used it for our initial WATM prototypes.
However, we note that any language that can be compiled to WebAssembly could be
used in the future, and that the resulting binaries can run in any
\water runtime build with WASI-compatible WebAssembly runtime.

\subsection{Security Consideration}

While WebAssembly provides substantial isolation between the transport module
and the runtime library, significantly mitigating the risk of attackers
executing malicious code, we note that WebAssembly is not inherently
impervious to binary-based attacks~\cite{wasm_security}. Also, despite the strict 
interface provided a strong restriction against any arbitrary malicious actions 
being performed on the host environment, we note that it is still possible for malicious 
WATMs to make arbitrary connections and potentially leak sensitive data from a
circumvention tool loading an arbitrary WATM. 
As with any software, it is important that we provide a path of trust, using things 
such as code signing and verification to ensure that the WATM that a 
client chooses to run is trusted by the deploying application. This is
also true of other circumvention tools working with pluggable elements, though
those are often integrated at compile time. For example, the Tor project packages 
and signs pluggable transports that are then launched with \texttt{execve} on the 
client. The WATMs used by \water clients are similar and should be packaged 
and signed by trusted parties (e.g. circumvention tool developers) before being loaded 
into a circumvention tool.

%% file: 4impl.tex
\section{Implementation}

\subsection{Runtime Library}
\label{sec:runtime-libray-impl}
We build \water runtime libraries in both Go and Rust to demonstrate the
cross-platform and cross-language abilities of our approach, with each 
providing a native-styled network programming interface for their respective 
programming language. To avoid excessive duplicated work in
implementing a new WebAssembly runtime and keep the design of \water 
runtime-independent, we only employ standard WebAssembly and WASI interfaces
which are available in every standard-complete WebAssembly runtime library~\cite{wasmtime, wazero, wasmer, wasmedge}. 
Currently, \water is built with \textit{wasmtime}~\cite{wasmtime} in Rust and 
with \textit{wazero}~\cite{wazero} in Go. In addition, we provide starter code, 
examples, and detailed documentation for developers to build their tool with \water. 
The open-source repository has been published on GitHub~\cite{water, watm-go, water-rs}.

\subsection{WebAssembly Transport Module (WATM)}
Currently, Rust is the only language with complete official native support for 
compilation to WebAssembly~+~WASI, with TinyGo as another very popular choice 
that allows the compilation of Go code for wasm-wasi. With the support of wasm-wasi being 
added to more programming languages, we are currently seeing at least 5 languages, 
including Rust, Go, Python, C, and Zig, becoming possible candidate languages for 
WATMs. 

\subsubsection{Provided Examples}
\label{sec:watm_examples}

A few example WATMs 
are provided by us to demonstrate the viability.

\textbf{plain.wasm} (available in Rust \& TinyGo) implements an \textit{identity} transform WATM that simply
copies over the bytes as-received, bi-directionally.

\textbf{reverse.wasm} (available in Rust \& TinyGo) reverses the bytes passed (e.g., from \texttt{ABCD} to
\texttt{DCBA}) before writing the result to the other end.

\textbf{shadowsocks.wasm} (available in Rust) demonstrates that WATMs can implement much more complex
protocols, such as Shadowsocks. Rather than mimicking Shadowsocks as prior work did, 
\water is able to build the real Shadowsocks client that works with an
unmodified server running \texttt{shadowsocks-rust}~\cite{shadowsocks-rust}
v1.17.0. To build shadowsocks.wasm, we started with the vanilla shadowsocks-rust
and identified 417 lines of code from a file that implements the core part of
Shadowsocks (i.e. cryptography and message formatting). We removed all
but the default features to minimize the code size and added 142(15\% of total) lines as 
the wrapper code to interface with our WATM specs (more details are provided in
Appendix~\ref{sec:shadowsocks-water}). 
Reusing the original codebase allows us to easily apply updates to our
shadowsocks.wasm based on upstream changes: we successfully applied the exact 
patch~\cite{ss-patch} for shadowsocks-rust defending against the China's blocking 
of fully-encrypted protocols~\cite{gfwreport-fully_encrypted} without any changes to the patch commit.

%% file: 5eval.tex
\section{Evaluation}

We evaluated our implementation in Rust, and compared latency and throughput with
Proteus~\cite{wails-proteus} and native (Rust) network performance. The evaluation was
conducted on the CloudLab testbed~\cite{Duplyakin+:ATC19} on \textit{c6525-25g}
(16-core AMD 7302P@3GHz, 128GB ECC RAM).

\subsection{Performance Metrics}

\TblLatencyThroughputCloudlabtopologyFinal

Our analysis of \water's performance is based on experiments for two different
transport protocols comparing \water, Proteus, and native implementations:
\textit{plain} and \textit{shadowsocks}. We compared the performance of each on a
set of writes using buffer sizes ranging from 1B to 4096B and noticed buffer size
can have a noticeable impact on latency and throughput. 
For baseline native
implementations we use raw TCP for \textit{plain}, and vanilla shadowsocks-rust
for \textit{shadowsocks}. In Proteus, we use a PSF implementing \textit{the identity
transform} for the \textit{plain} protocol, and the shadowsocks PSF implemented
from the original paper~\cite{wails-proteus} for the \textit{shadowsocks}.

Focusing on the 512-byte packet size, the \textit{shadowsocks} variant of \water
demonstrated throughput comparable to Proteus, but with improved latency. In the
\textit{plain} setup, \water closely matched the performance of native TCP
in both latency and throughput. Despite WebAssembly's virtualization introducing
a discernible overhead compared to native methods, we regard \water's performance as
highly promising.
Also, we have noticed it is possible to compile WebAssembly into native machine code 
with an Ahead-Of-Time (AOT) compiler~\cite{wazero} instead of executing it with an interpreter, 
which could mitigate such overhead.

The notable degradation in throughput for both \water and Proteus when
evaluating shadowsocks is in large part due to the cryptographic operations required. 
WebAssembly runtimes currently lack hardware acceleration for these operations, 
resulting in higher latency and lower throughput. However, support for hardware 
acceleration is being actively considered by the WebAssembly community~\cite{wasi-proposals}, 
and we anticipate that this will significantly improve \water's performance in the future. 
The performance of cryptographic operations in WebAssembly is examined further in Appendix~\ref{sec:crypto_performance}.

Lastly, we would mention that the testbed hosts were capable of saturating 
the 2~Gbps network link with the native transports, which is not a typical 
speed rate through any real-world Internet access provider. Thus, 
with shadowsocks-\water achieving a speed rate of around 50~Mbps, we believe 
it is very unlikely to be the bottleneck in a real-world circumvention scenario.
Further performance-based analysis can be found in Appendix~\ref{sec:more-benchmark-latency-throughput}.

%% file: 6discussion.tex
\section{Discussion}

\subsection{Advantages and Limitations}

\paragraph{Maximized code reuse}
Beyond the interchangeability of the WATMs, the use of WebAssembly also enables
existing tools (implemented in languages can be compiled to WASM) to easily
be converted into new WATMs. Appendix~\ref{sec:ss-code-change}
investigates this further, examining the code changes we made while porting an
existing circumvention tool to \water.

\paragraph{WebAssembly Limitations (Temporary)}
Given the limited official support for WASI in many programming languages, 
we have only proof-of-concept WATM implemented in Go and Rust available. 
However, we do see promising trends of programming languages embracing and 
adopting WASI/WASM compatibility. 

We also recognize that the use of WebAssembly introduces non-negligible overhead, due
in part to the lack of hardware acceleration support for cryptographic operations. 
However, WebAssembly is a rapidly evolving technology with a large and active community 
working to bring features like secure randomness sources, cryptographic acceleration, 
and network socket access into standards~\cite{wasi-proposals}. We expect that these 
features will only improve the potential of \water.

\subsection{Future Work}

Our current primary focus is to ensure that \water is a production ready
technology. We plan on working with several stakeholders to deploy \water to
real world users facilitating rapid and interoperable new circumvention
techniques.

\water could also assist in the discovery of new circumvention techniques. With
tools such as OONI~\cite{filasto2012ooni}, Censored
Planet~\cite{sundara2020censored}, and Ripe Atlas~\cite{staff2015ripe} the set
of probes available is rigid and limited by the software on the available
vantage points --- promoting a focus on \textit{WHAT} is blocked. Using \water
we could instead focus on \textit{HOW} network traffic is blocked by rapidly
iterating on probing experiments without redeploying entire applications.

Finally we intend to explore the use of Pseudo Random Functions (PRFs) in WATMs
such that the each WATM could have its own unique version of a transport
protocol. This would allow for the creation of a large number of unique
transports within a class making it more difficult for censors to block users
at scale.

We believe that this work only scratches the surface of the potential that
WebAssembly has to offer in the circumvention space. We hope that this work
inspires further exploration of the use of WebAssembly in circumvention tools.

\section{Conclusion}

In the paper we presented \water, a novel approach using WebAssembly to build 
circumvention transports that are rapidly deployable by lowering the barriers 
of deploying circumvention techniques, and
simplify the deployment cycle down to the delivery of a new binary file to
user's device. We use WebAssembly to provide a sandboxed environment for safely
running these binaries, and provide programmable libraries to facilitate integration
into existing circumvention tools and/or building new transport modules without 
knowledge about WebAssembly.

%% file: Appendix.tex
\appendix

\section{Cryptography Performance in WebAssembly}
\label{sec:crypto_performance}

WebAssembly is a new technology still in its early stage of development, and provides 
an isolated virtual execution environment like other virtual machines (VM). 
Therefore, it is not too surprising that it inherently lacks the support for native 
hardware acceleration for cryptographic operations (e.g. SIMD or AESNI). However, it is still 
important to evaluate and understand the performance of WebAssembly in terms of cryptographic
operations. We present results of AES operations within WebAssembly in Table~\ref{tab:crypto_performance}.
\begin{table}[ht]   
    \begin{tabular}{lcc}
    \toprule
    Configuration & Native & WebAssembly \\
    \midrule
    AES\_256\_GCM - 256B & 230 µs & 5300 µs \\
    AES\_256\_GCM - 1.09G & 200 s & 500 s \\
    CHACHA20\_POLY1305 - 256B & 320 µs & 5400 µs \\
    CHACHA20\_POLY1305 - 1.09G & 180 s & 200 s \\
    \bottomrule
    \end{tabular}
    \caption{\label{tab:crypto_performance} Crypto Performance - run on 2021 Macbook Pro M1 Max, 10-core CPU, 64GB RAM. 
    Rounded average for 5 trials.}
\end{table}

It is also worth noting that it is possible for an optimizing compiler of WebAssembly to 
allow efficient cryptographic features, and there is also an existing proposal named 
\texttt{wasi-crypto} that defines a set of WebAssembly-native APIs to import cryptographic
operations from host~\cite{wasi-crypto}.

\section{Implementing \texttt{shadowsocks.wasm}}
\label{sec:shadowsocks-water}
We have been developing two distinct versions of \texttt{shadowsocks.wasm}: one original 
and one patched version according to \textit{shadowsocks-rust} to counteract blocking and 
demonstrate the feasibility of circumventing GFW. 
Both versions of \texttt{shadowsocks.wasm} are designed to handle Shadowsocks' core functionalities, including encryption, decryption,
and packaging, within the \water environment.

\subsection{PoC version \texttt{shadowsocks.wasm}}
The PoC version utilizes the client and server implementation of the \textit{shadowsocks-rust} library, 
integrating \texttt{shadowsocks.wasm} to just run the main logic for the protocol, which is basically 
packaging code with \textit{shadowsocks-crypto}~\cite{shadowsocks-crypto}. The initial challenge 
we faced was the limitation of WASI being in its developmental phase, currently supporting 
only 32-bit targets for compilation. However, we found that a 32-bit usize integer suffices 
for the core functionalities of encryption, decryption, and packet transmission. We plan 
to continually update our runtime library to follow the advancements in WASI's latest standard.

\subsection{Porting from shadowsocks-rust}
\label{sec:ss-code-change}

\TblShadowsocksCodeCompare

We minimize the required changes to 
shadowsocks-rust's client code by identifying the protocol specification section 
(i.e. encryption, decryption, message framing) and reduced the feature support 
to only AEAD ciphers and direct connections(act like a transparent relay) for now, along with tunnel creation 
for asynchronous networking. We also had to implement a SOCKS5 listener to directly handle the incoming connections 
from external web browsers.
Table \ref{tab:ss-codechangecompare} showcases a segment of the comparative analysis 
between the core logic implementations in \textit{shadowsocks-rust} 
and \water-SS. The analysis indicates that, 
of the total 927 lines of code in \water-SS, 785 lines (approximately 
\textbf{85\%}) correspond with those in the 
official \textit{shadowsocks-rust} implementation. 
The remaining \textbf{142 lines} are primarily comprised of 
glue code, intentionally incorporated to incorporate all previously 
discussed enhancements. Notably, the amount of glue code should remain approximately 
consistent and does not proportionally increase with the expansion of the protocol implementation.

\subsection{Patching against GFW}
\label{sec:ss-patch}

The patch we applied was developed in response to China's move last year to block 
fully encrypted protocols, as reported by \cite{gfwreport-fully_encrypted}. This particular 
implementation, designed to mitigate the blocking of shadowsocks by the GFW, was proposed by 
gfw-report~\cite{ss-patch} and discussed in detail on Net4People~\cite{gfwreport-ss-discussion}. 
Our implementation of \texttt{shadowsocks.wasm} successfully incorporates this patch without necessitating 
any modification made to the patching commit.

Table~\ref{tab:ss-patch-compare} showcases the code comparison result of
the output of \texttt{diff} on the commit changes made while patching
the official \textit{shadowsocks-rust} and \water-shadowsocks. The specific commits compared
are the gfw-report \textit{shadowsocks-rust} patch commit~\cite{ss-patch} and the 
\water-shadowsocks patch commit~\cite{water-rs-patched-ss},
where it's obviously showing that the changes in \water is matching exactly the same changes in 
the official \textit{shadowsocks-rust} patch commit ignoring logging.

\TblSSPatchCodeDiff

\section{Latency and Throughput}
\label{sec:more-benchmark-latency-throughput}
\begin{figure}[ht]
    \centering
    \includegraphics[scale=0.5]{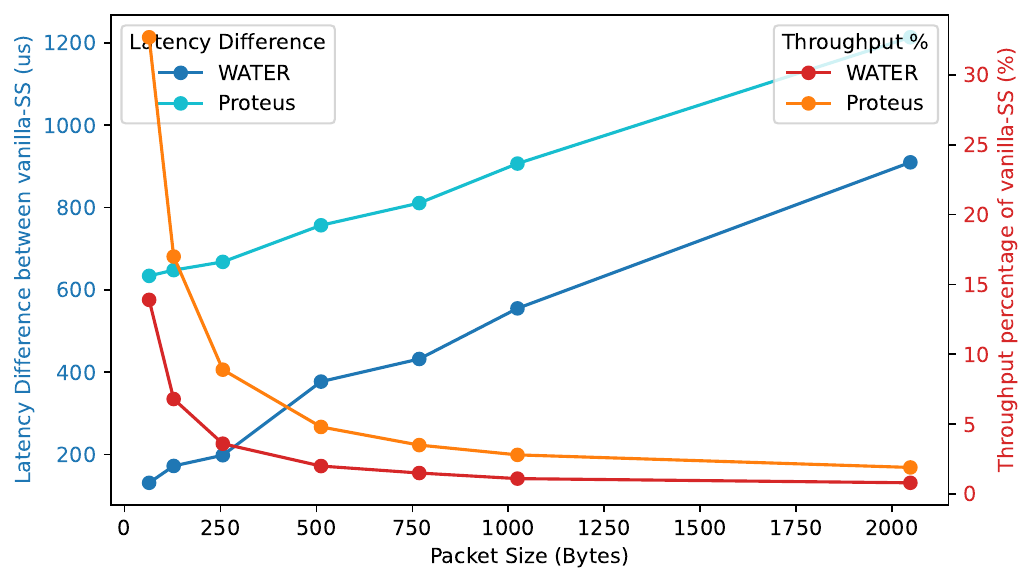}
    \caption{Latency \& Throughput Comparison with Vanilla-SS at Different Packet Sizes}
    \label{fig:latency_throughput_ss}
\end{figure}

We also present a detailed Table~\ref{tab:plain_latency_throughput_cloudlab_simulated}
outlining benchmark results for both latency and throughput across varying single packet sizes. 
In the table, raw TCP serves as the baseline for comparisons in the plain mode as we discussed in Section \ref{sec:watm_examples}.
Additionally, in Figure~\ref{fig:latency_throughput_ss}, we compared the latency and 
throughput of Shadowsocks implementations using \water and Proteus to provide a clearer 
picture of \water's performance and to identify the optimal packet size for 
balancing latency and throughput with \water.

\TblWaterPlainLatencyThroughputCompCloudlabtopology

\subsection{General Use Case performance}
\label{sec:apple-benchmark}
We conducted more real-world general use case tests on an Apple MacBook Pro 2021, 
equipped with a 10-core M1 Max CPU, 64GB of unified memory, and a 
32-core GPU. The results are presented in Table ~\ref{tab:ss_performance_macos}, 
showcasing performance metrics obtained using \texttt{iperf3} 
to connect from Michigan to a server in San Francisco.

\TblShadowsocksIperfMac

\section{WATER Workflow in detail}
\label{sec:workflow-detailed}

\begin{figure}[ht]
    \centering
    \includegraphics[width=\linewidth]{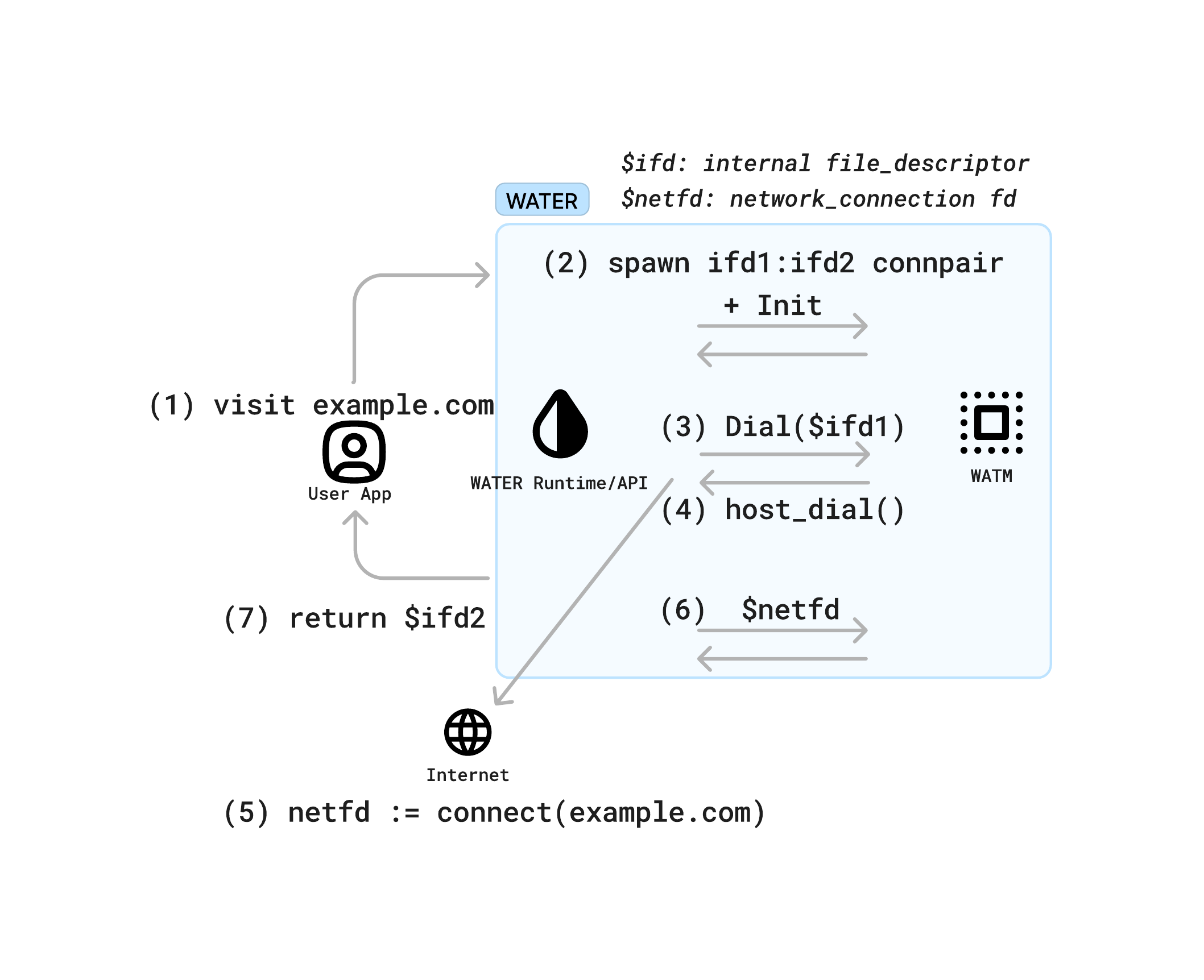}
    \caption{Step-by-step workflow of WATER when a new connection is dialed using WATER API by an integrating application.}
    \label{fig:workflow-detailed}
\end{figure}

In this appendix section, we provide a step-by-step workflow illustration of how 
\water establishes an outgoing connection in Figure \ref{fig:workflow-detailed}.

%% file: API.tex
\section{WATM usage in \water}
\label{sec:watm-api}

In this section we provide an up-to-date list of APIs we defined for WATM 
to interact with \water with WASI-based Imported/Exported functions.

\subsection{Imported Functions into WATM by \water Runtime}
This subsection outlines the functions imported from the \water Runtime to the WATM to allow WATM to interact 
with host-managed resources with restricted access. Each function is designed to facilitate specific operations within the WATM:

\begin{itemize}
    \item \textbf{\texttt{host\_dial()}}: A function that initiates a network connection, returning a network file descriptor (\texttt{net\_fd}).
    \item \textbf{\texttt{host\_accept()}}: A function designed to accept incoming network connections, similarly returning a network file descriptor (\texttt{net\_fd}).
    \item \textbf{\texttt{pull\_config()}}: This function retrieves the configuration settings, issuing a configuration file descriptor (\texttt{conf\_fd}).
\end{itemize}

\subsection{Exported WATM APIs}
This subsection describes WATM API functions exported by each WATM, detailing their purposes and return values to elucidate their roles:

\begin{itemize}
    \item \textbf{\texttt{init()}}: Initializes the WATM module, returning an error number (\texttt{errno}) as a signed 32-bit integer (\texttt{s32}) to indicate success (\texttt{0}) or failure.
    
    \item \textbf{\texttt{dial(internal\_fd)}}: Used in Dialer role, which establishes a network connection using an internal file descriptor, and returns a network file descriptor (\texttt{net\_fd}) as \texttt{s32}.
    
    \item \textbf{\texttt{accept(internal\_fd)}}: Used in Listener role, which accepts an incoming connection on an internal file descriptor, returning a network file descriptor (\texttt{net\_fd}) as \texttt{s32}.
    
    \item \textbf{\texttt{associate()}}: Used in Relay role, which associates an incoming connection with an outgoing connection, typically returning an error number (\texttt{errno}) as \texttt{s32} to indicate the outcome.
    
    \item \textbf{\texttt{worker()}}: Launches a worker thread and works as the assigned role, returning an error number (\texttt{errno}) as \texttt{s32}.
\end{itemize}

\subsection {WebAssembly System Interface (WASI)}
Besides the above mentioned imported/exported functions, our current WATM spec is based on WebAssembly System Interface 0.1.0 (a.k.a., WASI Preview 1 or 
\texttt{wasi\_snapshot\_preview1}. A WATM will expect all imports defined by WASI Preview 1 to be made accessible, 
which SHOULD always be the case out-of-box from any WASI-compliant WebAssembly runtime environment including the ones we 
mentioned in Section \ref{sec:runtime-libray-impl}.